\documentclass[10pt]{iopart}
\usepackage[numbers,sort&compress]{natbib}
\usepackage{graphicx}% Include figure files
\usepackage{dcolumn}% Align table columns on decimal point
\usepackage{bm}% bold math
\usepackage{color}
\usepackage{stmaryrd, xparse, mleftright}
\usepackage{array}
\usepackage{ulem}

\usepackage{blindtext, xcolor}
\usepackage{xcolor}
\usepackage{hyperref}
\hypersetup{
    colorlinks,%
    citecolor=blue,%
    linkcolor=blue,%
    urlcolor=blue
}

\begin{document}

\title{Noncentrosymmetric two-dimensional Weyl semimetals in porous Si/Ge structures}

\author{Emmanuel V. C. Lopes$^1$\footnote{\ead{emmanuel.lopes@ufu.br}}, Rogério J. Baierle$^2$, Roberto H. Miwa$^1$ and Tome M. Schmidt$^1$\footnote{\ead{tschmidt@ufu.br}}}
\address{%
 $^1$ Instituto de Física, Universidade Federal de Uberlândia, Uberlândia, Minas Gerais 38400-902, Brazil
}%
\address{%
 $^2$ Departamento de Física, Universidade Federal de Santa Maria, Santa Maria, Rio Grande do Sul 97105-990, Brazil
}%

\begin{abstract}
In this work we predict a family of noncentrosymmetric two-dimensional (2D) Weyl semimetals composed by porous Ge and SiGe structures. These systems are energetically stable graphenylene-like structures with a buckling, spontaneously breaking the inversion symmetry. The nontrivial topological phase for these 2D systems occurs just below the Fermi level, resulting in nonvanishing Berry curvature around the Weyl nodes. The emerged Weyl semimetals are protected by $C_3$ symmetry, presenting one-dimensional edge Fermi-arcs connecting Weyl points with opposite chiralities. Our findings complete the family of Weyl in condensed-matter physics, by predicting the first noncentrosymmetric class of 2D Weyl semimetals.
\end{abstract}

\maketitle
\ioptwocol

\section{Introduction}
Among the materials with topological nontrivial phase, the Weyl semimetals (WSMs) are the last predicted \cite{Murakami2008,Wan2011,Burkov2011,Yang2011} and experimentally observed ones \cite{ Lv2015,Xu2015,Xu2015TaAs,Huang2015,Weng2015, Lu2015,Xu2015Niob,Xu2016}. This class of materials attracts attention by their interesting properties, such as chiral anomaly and negative magnetoresistance~\cite{Son2013,Zyuzin2012}. They are potentially for applications in thermoelectricity, electronics, as well they present a connection with superconductivity, a subject in discussion~\cite{Sukhachov2020, Han2020, Arjona2019, Fu2020, Singh2021}. WSMs are characterized by massless states in the bulk of the material with nonvanishing Chern number~\cite{Yang2016}. They are in pairs of opposite chirality nodes, where each crossing point is a singularity of the Berry curvature, presenting a monopole chiral charge~\cite{Mele2018}. The opposite chiral nodes in three-dimensional (3D) system present a bulk-boundary correspondence resulting in surface Fermi-arcs~\cite{Lv2021}. The 3D WSMs has been observed in many materials, where the Weyl nodes arise under breaking one of the fundamental symmetries, inversion~\cite{Lv2015,Xu2015,Xu2015TaAs,Huang2015,Weng2015, Lu2015,Xu2015Niob,Xu2016} or time-reversal~\cite{Wan2011,Yang2011,Ruan2016,Chang2018,Liu_2019}.

With the success and potential application of 3D WSMs, one should ask if a Weyl point would survive in a two-dimensional (2D) system. The possibility of 2D WSMs has been investigated using theoretical models~\cite{Yang2016,Guo2019, Isobe2016, Hao2016}. The fulfillment of 2D WSMs would be more promising for quantum computing and storage nanodevices due to the 2D architecture~\cite{Bader2010, Li2016, Awschalom2007}. In 2D materials the Weyl points require extra symmetry protection, not needed in the counterpart 3D systems~\cite{Yang2016}. Recently it has been proposed WSMs in 2D systems such as MnNBr, Cr$_2$C, NiCS$_3$, CrN, VI$_3$, Mn$_2$NF$_2$, PtCl$_3$ and NpAs monolayers~\cite{Shi2021,Meng2021, Li2021, He2020, Jia2020, Wei2022, You2019, Zou_2021}, all of them belonging to time-reversal symmetry breaking systems. For noncentrosymmetric systems, only quadratic Weyl semimetals have been proposed in a 2D heterostructure~\cite{Zhao2022}.

In this work, we predict noncentrosymmetric 2D WSMs protected by crystal symmetry in porous structures. The systems are similar to the graphenylene \cite{Song2013, Du2017, galvao2012} but composed by Ge or SiGe. While graphenylene is planar sp$^2$-carbon allotrope, the porous Ge and SiGe present a buckling, spontaneously breaking the inversion symmetry. The presence of topological nontrivial character just below the Fermi level, and the absence of inversion symmetry, are basic conditions to emerge pairs of Weyl nodes. Our results show that the Weyl nodes are protected by $C_3$ crystal symmetry, assuring their stability. The bulk-boundary correspondence is mediated by large energy dispersion edge arcs, connecting Weyl points with opposite chirality.

\section{Methodology}
Our calculations was based on first-principles within the density functional theory using the projector augmented waves method, implemented in Vienna {\it ab initio} simulation package (VASP)~\cite{Kresse1993, Kresse1996}. The exchange-correlation term was described using the generalized gradient approximation (GGA) in Perdiew-Burke-Ernzehof's (PBE) functional~\cite{Perdew1996}, including fully relativistic pseudopotentials. The plane wave basis set was used with a kinect energy cut-off of 400~eV. To map the Brillouin zone it was used the Monkhost-Pack k-mesh grid of $11~\times~11~\times~1$. The topological properties such as Weyl chirality, Berry curvature, and projected edge states have been computed using the WannierTools package~\cite{Wu2017}.

\section{Results and discussion}
\begin{figure}[htb]
    \centering
    \includegraphics[width=8cm, height =6cm]{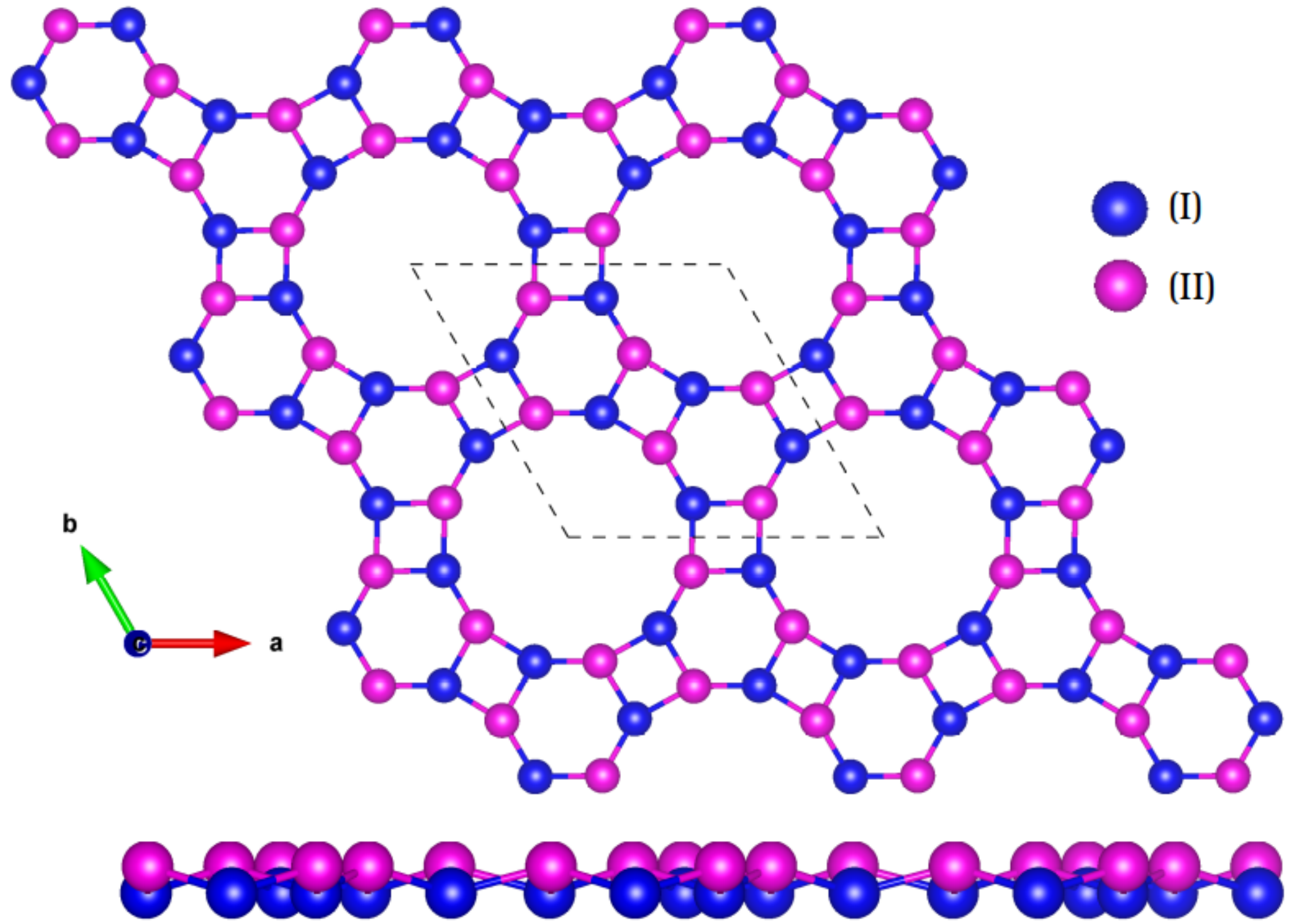}
     \caption{Top and side view of the crystal structure, showing the unit cell in dashed lines. Atoms (I) and (II) can be both Ge, or (I) Si and (II) Ge. Atom (I) is at lower position as compared to atom (II).}
     \label{crystal}
\end{figure}

The porous layered structures studied here are formed by squares connected to hexagons with large pores as shown in Fig.~\ref{crystal}. The blue and magenta atoms can be either Si, Ge or both elements. All layers containing Si and/or Ge present a buckling degree, as shown in the figure, distinct from the centrosymmetric sp$^2$-carbon graphenylene, which is planar~\cite{Song2013, Du2017, Kochaev2020}. The stability of these structures was confirmed through phonon calculations and \textit{ab initio} molecular dynamics simulations at finite temperatures~\cite{Kremer2023}. Also the pores diameter of the Si/Ge structures are quite large, around 9~{\AA}, as compared to the graphenylene porous around 6~{\AA}~\cite{Du2017}, open it for new functionalities. The fully optimized lattice parameters were  $a = 10.60$, $10.92$ and $11.17$~{\AA} for Si, SiGe and Ge porous layers, respectively.

\begin{figure}[htb]
    \centering
    \includegraphics[width=8.3cm, height =5.0cm]{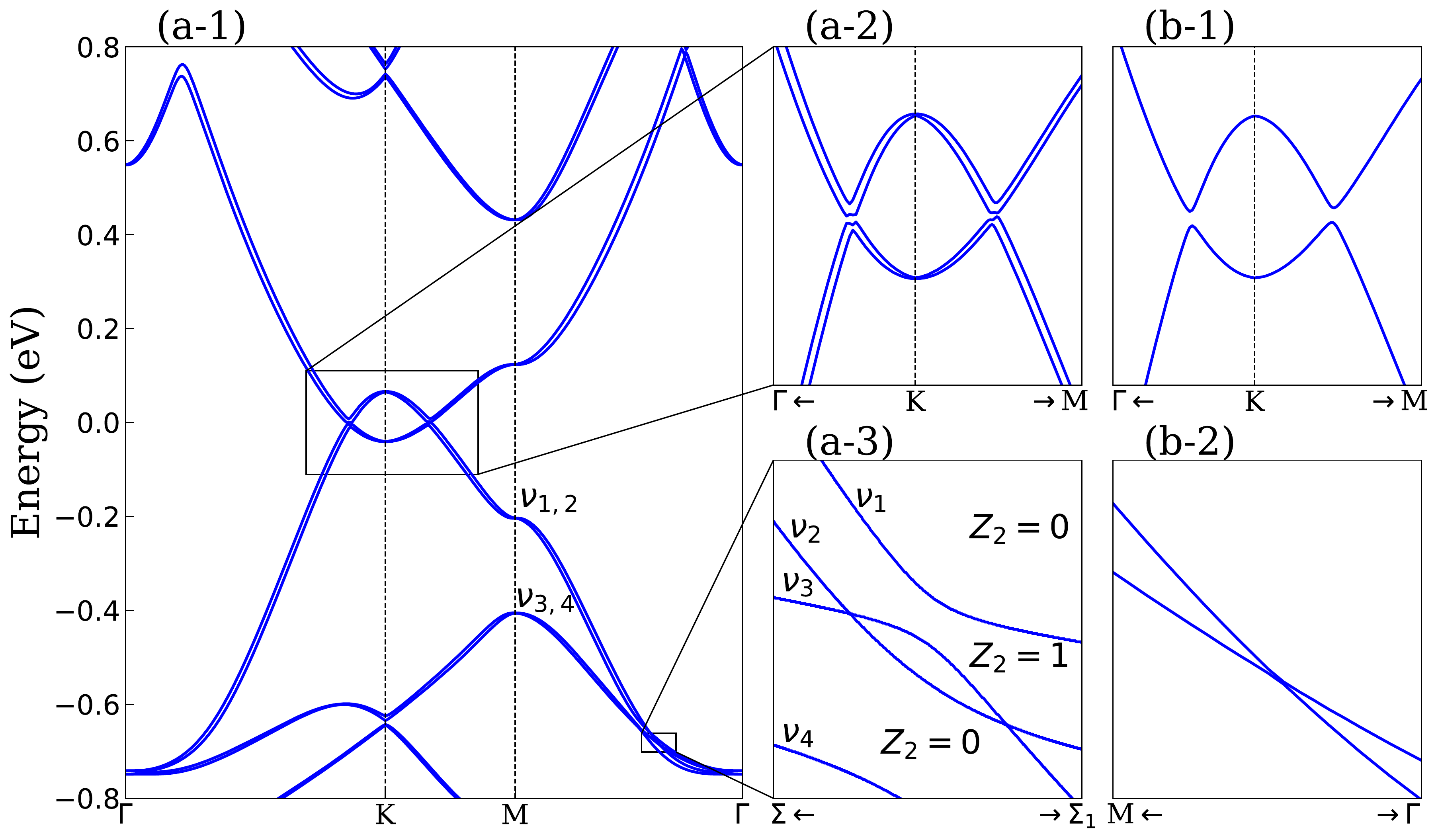}
    \includegraphics[width=8.3cm, height =5.0cm]{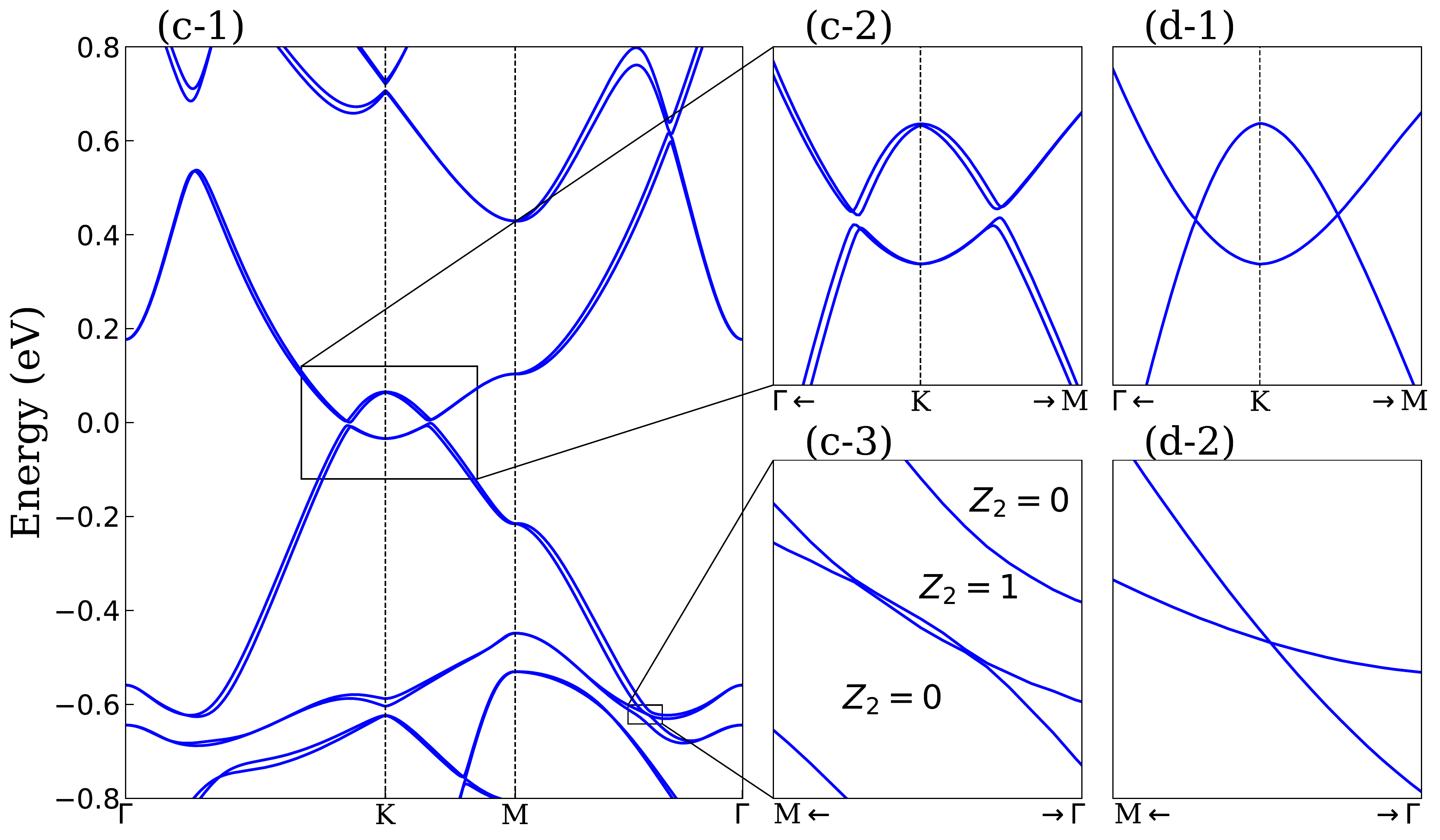}
     \caption{Band structure for porous SiGe (a)-(b) and Ge (c)-(d). The middle column are enlarged bands around the Fermi level and at the Weyl points. The last column are the same as the middle one, but without SOC.}
     \label{Bands}
\end{figure}

 Our first-principles results shows that the porous structure of Si is metallic (not shown here). The SiGe and Ge structures present narrow band gap, as can be seen in Fig.~\ref{Bands}. The band gap of the Ge comes from the spin-orbit coupling (SOC), without SOC it is metallic (Figs.~\ref{Bands}(c-1) and~\ref{Bands}(d-1)). A gap opening due to SOC is typically from topological nontrivial phase, as will be discussed below. The gapless for Si structure is due to its low SOC ($\lambda_{Si}=3.9$~meV compared to $\lambda_{Ge}=43$~meV~\cite{Liu2011}). For porous SiGe structure, due to the reduced symmetry, there is a band gap even without SOC, and the SOC effect splits the bands as can be seen in Figs.~\ref{Bands}(a-2) and \ref{Bands}(b-1).

The top of valence band at the $\Gamma$ point for SiGe has two pairs of bands labeled $\nu_{1,2}$ and $\nu_{3,4}$ in Fig.~\ref{Bands}(a-1). The effective mass at the $\Gamma$ point will be described by electrons instead of holes, resembling $\pi-\pi^*$ graphene bands, that has a negative band gap at the $\Gamma$ point. However, here we have two pairs of bands that will change the topological phase. We use the Wannier charge center method \cite{Soluyanov2011} to compute the topological invariants for the porous noncentrosymmetric system.
The computed Z$_2$ index at the Fermi level (starting from $\nu_1$ band) shows a trivial system. However starting from $\nu_2$ band the system becomes topologically nontrivial (Z$_2=1$), leading to a couple of interesting band-crossings, as can be seen in Fig.~\ref{Bands}(a-3). Similar results are obtained for Ge structure as shown in Fig.~\ref{Bands}(c-3).
Without SOC the bands are double degenerate, with massless Dirac-like crossings fourfold degenerated (Figs.~\ref{Bands}(b-2) and \ref{Bands}(d-2)). The SOC split up the degeneracy, resulting pairs of nondegenerate bands crossing each other at two specific points. This is typically of Weyl semimetals, since the systems are noncentrosymmetric, but preserving time-reversal symmetry, as is required for a nonvanishing Berry curvature.
For Si porous structure the low value of SOC parameter does not result in Weyl nodes.

\begin{figure}[htb]
    \centering
    \includegraphics[width=8.5cm, height =4.0cm]{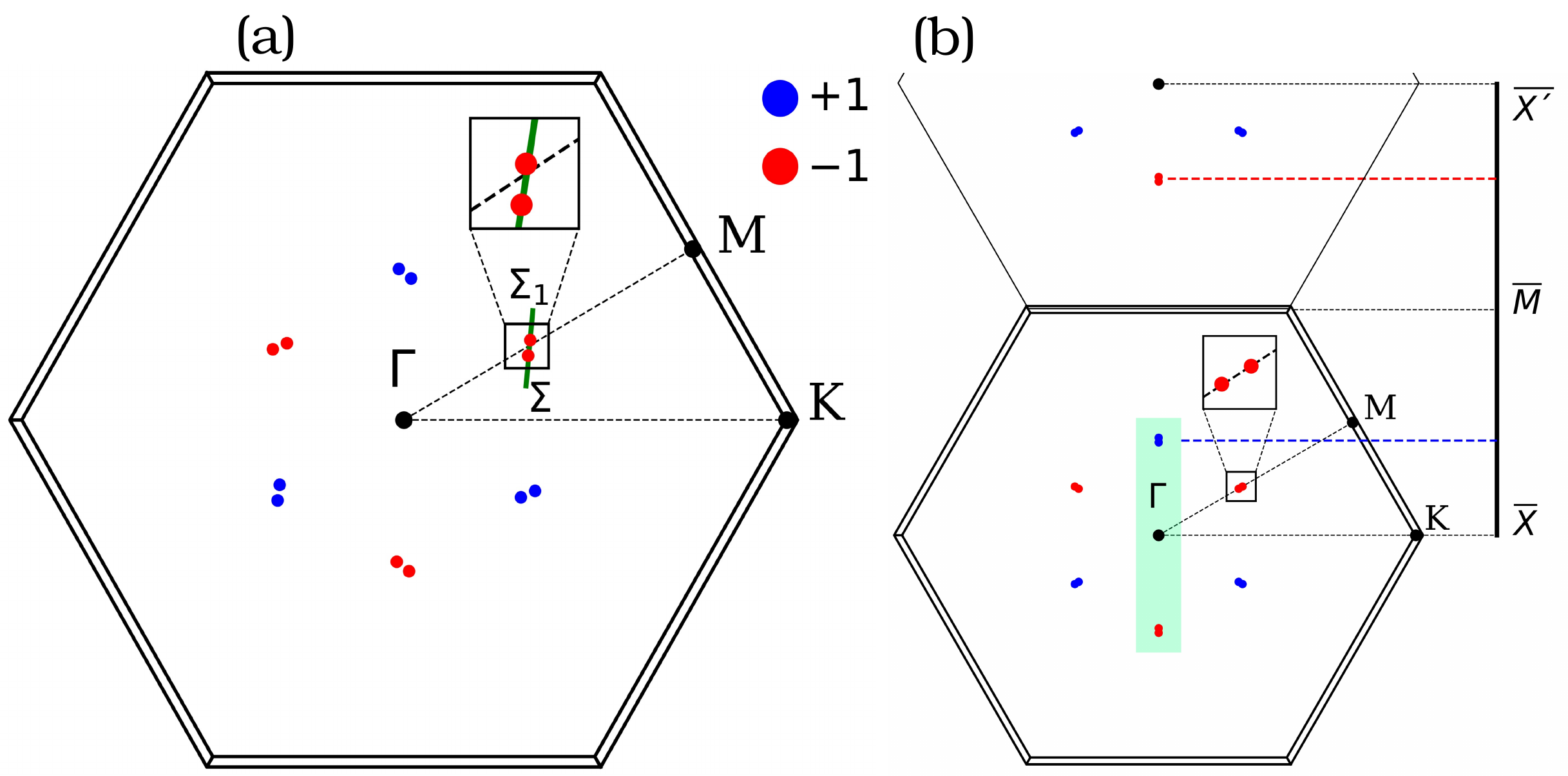} 
     \caption{Weyl point chiralities on the 2D Brillouin zone for porous SiGe (a) and Ge (b). Inset of each figure is a zoom around the nodes to illustrate the splitting of Weyl pairs with the same chirality. The red circles corresponds to negative chirality and blue circles positive chirality. }
     \label{chirality}
\end{figure}

In order to find the topological characteristic of a Dirac crossing, we compute the topological charge around each crossing point. Starting from the Weyl equation \cite{Weyl1929}, in~\ref{Appendix} we derive an expression for the 2D Weyl chirality ($\chi^{2D}$)
\begin{equation}
\label{Chirality}
\chi^{2D} = \frac{1}{\pi}\oint_{_{l}}\boldsymbol{A}(\boldsymbol{k})\cdot d\boldsymbol{k}.
\end{equation}
Here $\boldsymbol{A}(\boldsymbol{k})$ corresponds to the Berry connection, and $l$ is a closed loop in momentum space around the 2D Weyl point. If nontrivial topological phase is present, each $\chi^{2D}$ carries $\pm 1$ chirality value. By using the equation above our results shows that each crossing pair has the same chirality, but the all system is neutral, as can be seen in Fig.~\ref{chirality}, in accordance with fermion doubling theorem~\cite{Nielsen1983, Nielsen1981}. Weyl pairs with the same chirality are less common, but have been reported in 3D materials as well, such as magnetic-doped Sn$_{1-x}$Pb$_x$(Se,Te), AgBi(Cr$_2$O$_7$)$_2$~\cite{Liu_2019, Chang2018} and, in the 2D Cr$_2$C monolayer~\cite{Meng2021}. Unlike in 3D time-reversal symmetry protected WSMs, where the Weyl points present the same chirality at $\boldsymbol{k}$ and $-\boldsymbol{k}$, its 2D counterpart will present different chiralities~\cite{Zhao2022}. By applying time-reversal operation at the Berry curvature, Eq.~\ref{Bcurv}, it will change the sign, consequently changing the chirality, $\chi^{2D}(\boldsymbol{k})=-\chi^{2D}(-\boldsymbol{k})$.

As Ge and SiGe structures belong to a distinct space group symmetries, $P_{622}$ (or $D_{6}^1$) and $P_{6}$ (or $C_{6}^1$), respectively, the 2D Weyl nodes are located at similar positions but at different directions. The pair of Weyl nodes in germanylene are along the high-symmetry $\Gamma-$M direction (Fig.~\ref{chirality}(b)), located at $(0.216,0.0,0.0)$ and $(0.206,0.0,0.0)$ positions in units of $2\pi/a$ (where $a$ is the lattice parameter). For the low symmetric SiGe, the Weyl crossings are just out of the $\Gamma-$M direction, in $\Sigma-\Sigma_1$ path, represented by the green line in Fig.~\ref{chirality}(a). The crossings are located at $(0.188,0.0075,0.0)$ and $(0.185,-0.0103,0.0)$ positions at the same unit above. Both systems present a total of six pairs of Weyl crossings protected by crystal symmetry, as is required for 2D WSMs \cite{Yang2016}, here the $C_{3}$ rotation symmetry.

\begin{figure}[htb]
    \centering
    \includegraphics[width=8.3cm, height =4.0cm]{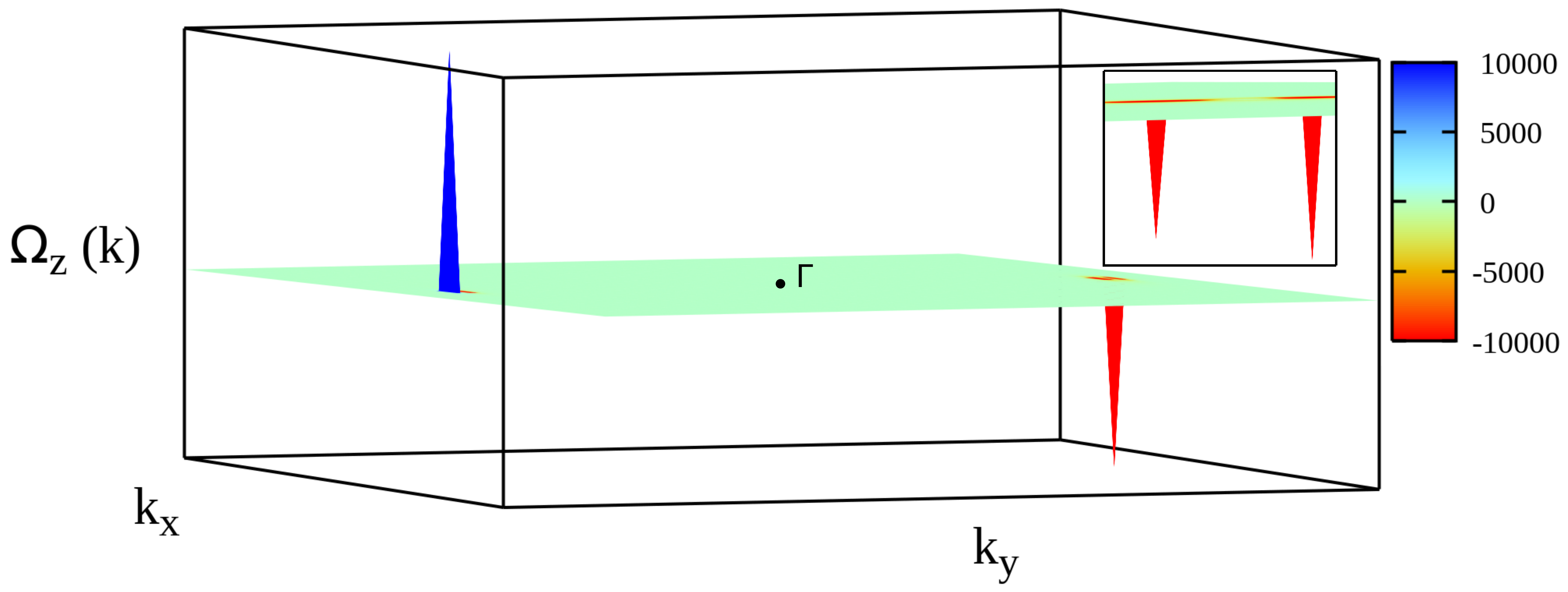}
     \caption{Computed Berry curvature for porous Ge structure. The k$_{\rm x}$k$_{\rm y}$-plane corresponds to the aquamarine rectangle  marked in Fig.~\ref{chirality}(b). The inset is a zoom showing the split of Weyl pairs with negative chirality.}
     \label{Berrycurvature} 
\end{figure}

The Berry curvature for a 2D Weyl system, as deduced in~\ref{Appendix}, is aligned perpendicular to the $xy$-plane formed by a delta function centered at the Dirac crossing point $k_0$ of the Brillouin zone (BZ)
\begin{equation}\label{BerryC}
    \boldsymbol{\Omega}_{k_0}^{2D}=\pi\delta^2(\boldsymbol{k}-\boldsymbol{k_0})\boldsymbol{e}_{kz}.
\end{equation}
We compute the Berry curvature for germanylene using the DFT results projected onto the Wannier orbital basis. Fig.~\ref{Berrycurvature} shows the Berry curvature for a region in the BZ containing two pairs of Weyl points (rectangular dashed region marked in Fig.~\ref{chirality}(b)). The computed Berry curvature from our first-principles calculation is a didactic result, showing delta functions centered at each Weyl point (inset of Fig.~\ref{Berrycurvature} is a better resolution, showing the splitting of $\chi^{2D}=-1$ chirality pair). These results are in concordance with the Berry curvature property for time-reversal symmetry protected crystals $\mathbf{\Omega(k)}=-\mathbf{\Omega(-k)}$~\cite{Yang2016}. 

\begin{figure}[htb]
    \centering
    \includegraphics[width=8.3cm, height = 7cm]{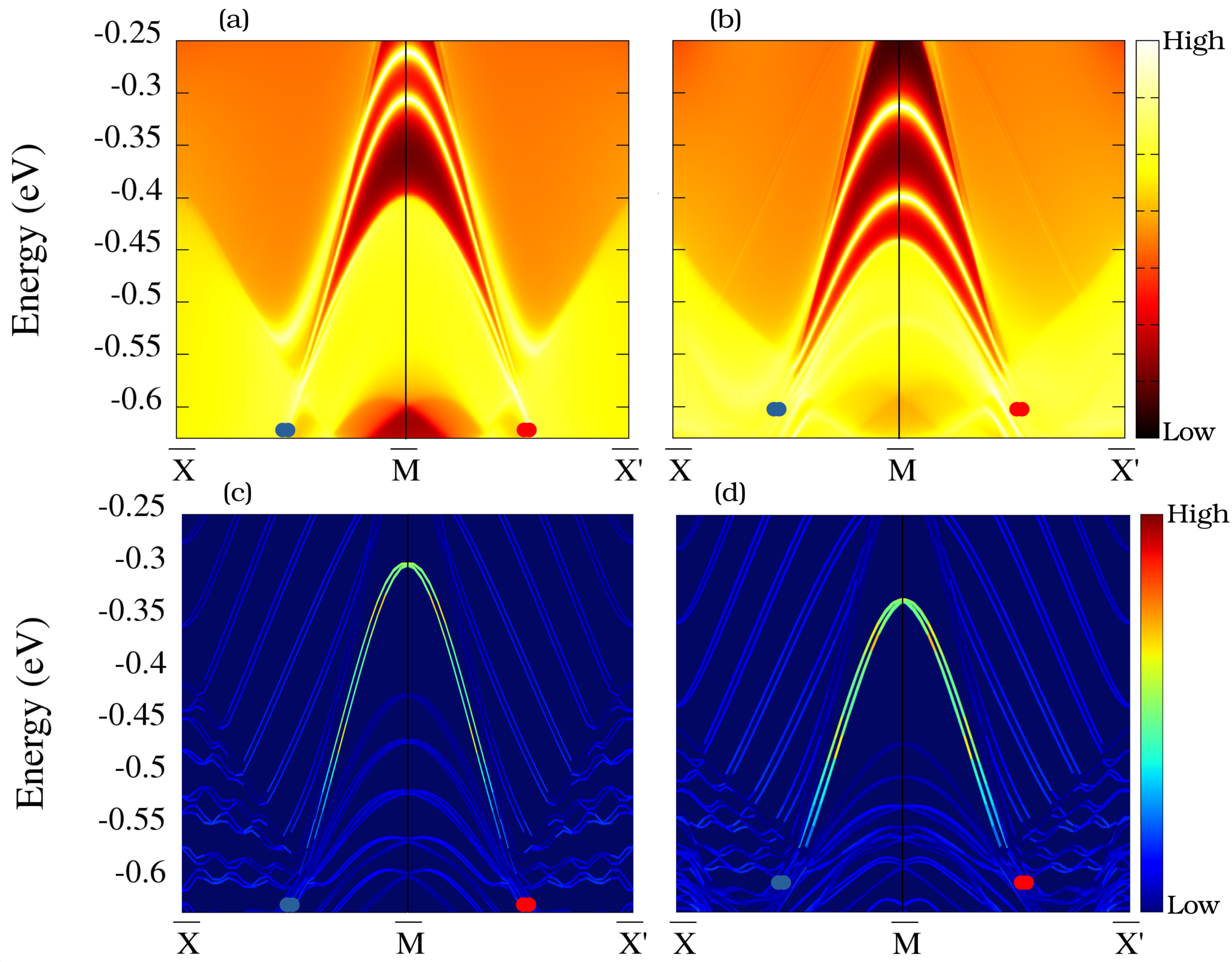}  
     \caption{Projected 1D edge states connecting Weyl points with opposite chirality for porous SiGe (a) and Ge (b) ribbons, using a Wannier basis scheme. In (c) and (d) are the same projections with hydrogen saturated edge states, using first-principles calculations. The edge $\overline{\rm X}-\overline{\rm M}-\overline{\rm X'}$ path is from 0 to $+2\pi/a$, as ilustrated in Fig.~\ref{chirality}(b).}
     \label{ARPES} 
\end{figure}

One of the main characteristics of topological materials is the bulk-boundary correspondence. Topological states of a TI connect the valence to conduction band, while in WSMs, this correspondence is expressed by surface states that connect Weyl points with opposite chiralities, the Fermi-arcs. In 2D materials, the topological states of a TI are massless states at the edges, while the presence of Weyl nodes inside the 2D bulk induces open curves restricted to just one momentum direction. These edge Fermi-arcs can be visualized by performing a band structure calculation of a ribbon. Using Wannier basis functions obtained from the first-principles bulk calculation, we show in Figs.~\ref{ARPES}(a) and \ref{ARPES}(b) the edge bands projected from $\Gamma$ to $\Gamma + 2\pi/a$, connecting Weyl nodes with opposite chiralities for SiGe and Ge porous structures, respectively. There are two edge states connecting two pairs of Weyl nodes, projected along a k-line perpendicular to the $\Gamma$-K direction (see Fig.~\ref{chirality}(b)). It is worth to note that the arcs are not degenerate at the time-reversal invariant momentum $\overline{M}$ point. As the ribbon edge is a rigid cut of the bulk, dangling bonds emerge leading to local magnetic moments due to unpaired electrons, opening up the Kramers degeneracy of the edge states. It is known that the Fermi-arcs are not topologically protected against weak disorder, as well they are affected by different types of perturbations, such as bulk-boundary hybridization and dangling bonds~\cite{parkin2021,yan2015,murakami2014, sarma2018}. In order to avoid the dangling bonds effects, we performed a first-principles calculation with a ribbon saturated at the edge with hydrogen atoms. In this passivated system, the local magnetic moments vanish. As we can see in Figs.~\ref{ARPES}(c) and \ref{ARPES}(d) the edge arcs are now merged at the $\overline{M}$ point, recovering the whole system symmetry, preserving the Kramers degeneracy.
The spin of the Fermi-arc bands is lifted, however there is no net magnetization in the system, preserving time-reversal symmetry. It can be note that the edge arcs present large energy dispersions, a characteristic also observed in 2D WSMs where the time-reversal symmetry is broken~\cite{Meng2021, Wei2022, Shi2021}.

\section{Conclusion}
In summary, based on first-principles results and topological invariant calculations, we propose a family of noncentrosymmetric 2D WSMs. It is a porous structure, similar to the graphenylene, but composed by Ge or SiGe. The structure presents a buckling, spontaneously breaking the inversion symmetry. By computing the Weyl chirality in both systems, we identify six pairs of Weyl nodes protected by $C_3$ crystal symmetry. The computed 1D edge Fermi-arc shows large energy dispersion connecting Weyl points with opposite chiralities. Our findings complete the family of Weyls in condensed-matter physics, by predicting the first noncentrosymmetric class of 2D WSMs.

\section*{Note}

During the revision process to submit this article, an arXiv preprint~\cite{Lu2023} was uploaded, reporting the first experimental verification of a 2D Weyl semimetal observed in epitaxial bismuthene. Those findings corroborate our results, demonstrating the feasibility of 2D WSMs.

\section*{Acknowledgments}

The autors acknowledge brazilian agencies FAPEMIG, CAPES, CNPq and INCT-Nanocarbono for financial support. Also Laboratório Nacional de Computação Científica (LNCC-SCAFMat2) and CENAPAD-SP for the computational resources.

\appendix
\section{Two-dimensional Weyl Chirality}~\label{Appendix}
The Hamiltonian proposed by Weyl~\cite{Weyl1929} can be used for crossing points in condensed matter physics, by including a velocity  $v_F$ for the fermions in the form~\cite{Burkov2018}
\begin{equation}
    H={\hbar}v_F\boldsymbol{k\cdot\sigma}.
\end{equation}
Here $\boldsymbol{\sigma}$ corresponds to the Pauli's matrix and $\boldsymbol{k}$ the crystal momentum. The isotropic 2D Weyl Hamiltonian in the $xy$-plane can be written as
\begin{equation}\label{3}
    H={\hbar}v_F(k_x\sigma_x+k_y\sigma_y).
\end{equation}
The solution of Schrödinger's equation for the Eq.~\ref{3} Hamiltonian, gives the eigenvalues $\epsilon=\pm~{\hbar}v_Fk$, with $k=\sqrt{(k_{x}^2+k_{y}^2)}$. Using the more appropriate cylindrical coordinates, the respective eigenfunctions are given by
\begin{equation}
    \psi_{\pm{\hbar v_{F}k}} = \frac{1}{\sqrt{2}} \left(
    \begin{array}{c}
        \pm e^{-i\theta} \\
        1
    \end{array}
    \right),
\end{equation}
with positive (negative) exponential term for positive (negative) momentum eigenvalue.

In order to compute a topological variable associated to a Weyl point, like a chiral charge, we start from the Berry phase~\cite{Berry1984}. The corresponding Berry connection can be computed from the eigenfunctions above
\begin{equation}
    \boldsymbol{A}(\boldsymbol{k})=i\langle\psi_{\pm}\vert\nabla_{\boldsymbol{k}}~\vert\psi_{\pm}\rangle,
\end{equation}
where $\nabla_{\boldsymbol{k}}$ is the gradient operator. The Berry connection for each momentum eigenvalue is then given by 
\begin{equation}\label{Berryconnection}
    \boldsymbol{A}(\boldsymbol{k})={\pm}\frac{1}{2k}\boldsymbol{e}_{\theta}.
\end{equation}

For 3D systems the chiral topological charge is computed by the Berry flux over a 2D closed Fermi surface around the Weyl node. While for a 2D system, the Berry flux can be obtain by integrated over a closed Fermi ring around the Weyl node
\begin{equation}
    \varphi=\oint_{l}\boldsymbol{A}(\boldsymbol{k})\cdot d\boldsymbol{k}.
\end{equation}
By using the derived Berry connection (Eq.~\ref{Berryconnection}), the integral gives a nonvanishing flux  $\varphi~=\pm\pi$. In this way we can define a topological variable associated with the Weyl chirality of a 2D system, by the expression
\begin{equation}\label{Chiral}
    \chi=\frac{1}{\pi}\oint_{l}\boldsymbol{A}(\boldsymbol{k})\cdot d\boldsymbol{k}=\pm1.
\end{equation}

 Another characteristic of Weyl physics can be derived from the Berry curvature, which is given by the curl of Berry connection, $\boldsymbol{\Omega}=\nabla\times\boldsymbol{A}$. It can be notted, by taken the curl in cylindrical coordinates, that the Berry curvature is null for $k\neq0$. In the limit that $\boldsymbol{k}$ tends to zero, we have a divergence, which means that, in this case, the Berry curvature is a delta function
\begin{equation}\label{Bcurv}
    \boldsymbol{\Omega}=\pm\pi\delta^2(\boldsymbol{k})\boldsymbol{e}_{kz}.
\end{equation}
Applying the Stokes' theorem to the Berry curvature of Eq.~\ref{Bcurv}, integrating over the 2D BZ, results in $\chi=\pm 1$, the same as Eq.~\ref{Chiral} for the chirality of 2D Weyls.

\bibliographystyle{iopart-num}

\bibliography{references.bib}

\providecommand{\newblock}{}
\begin{thebibliography}{10}
\expandafter\ifx\csname url\endcsname\relax
  \def\url#1{{\tt #1}}\fi
\expandafter\ifx\csname urlprefix\endcsname\relax\def\urlprefix{URL }\fi
\providecommand{\eprint}[2][]{\url{#2}}

\bibitem{Murakami2008}
Murakami S 2008 {\em New Journal of Physics\/} {\bf 10} 029802

\bibitem{Wan2011}
Wan X, Turner A~M, Vishwanath A and Savrasov S~Y 2011 {\em Phys. Rev. B\/} {\bf 83}(20) 205101

\bibitem{Burkov2011}
Burkov A~A and Balents L 2011 {\em Phys. Rev. Lett.\/} {\bf 107}(12) 127205

\bibitem{Yang2011}
Yang K~Y, Lu Y~M and Ran Y 2011 {\em Phys. Rev. B\/} {\bf 84}(7) 075129

\bibitem{Lv2015}
Lv B~Q, Xu N, Weng H~M, Ma J~Z, Richard P, Huang X~C, Zhao L~X, Chen G~F, Matt C~E, Bisti F, Strocov V~N, Mesot J, Fang Z, Dai X, Qian T, Shi M and Ding H 2015 {\em Nature Physics\/} {\bf 11} 724--727

\bibitem{Xu2015}
Xu S~Y, Belopolski I, Sanchez D~S, Zhang C, Chang G, Guo C, Bian G, Yuan Z, Lu H, Chang T~R, Shibayev P~P, Prokopovych M~L, Alidoust N, Zheng H, Lee C~C, Huang S~M, Sankar R, Chou F, Hsu C~H, Jeng H~T, Bansil A, Neupert T, Strocov V~N, Lin H, Jia S and Hasan M~Z 2015 {\em Science Advances\/} {\bf 1} e1501092

\bibitem{Xu2015TaAs}
Xu S~Y, Belopolski I, Alidoust N, Neupane M, Bian G, Zhang C, Sankar R, Chang G, Yuan Z, Lee C~C, Huang S~M, Zheng H, Ma J, Sanchez D~S, Wang B, Bansil A, Chou F, Shibayev P~P, Lin H, Jia S and Hasan M~Z 2015 {\em Science\/} {\bf 349} 613--617

\bibitem{Huang2015}
Huang S~M, Xu S~Y, Belopolski I, Lee C~C, Chang G, Wang B, Alidoust N, Bian G, Neupane M, Zhang C, Jia S, Bansil A, Lin H and Hasan M~Z 2015 {\em Nature Communications\/} {\bf 6}

\bibitem{Weng2015}
Weng H, Fang C, Fang Z, Bernevig B~A and Dai X 2015 {\em Phys. Rev. X\/} {\bf 5}(1) 011029

\bibitem{Lu2015}
Lu L, Wang Z, Ye D, Ran L, Fu L, Joannopoulos J~D and Soljačić M 2015 {\em Science\/} {\bf 349} 622--624

\bibitem{Xu2015Niob}
Xu S~Y, Alidoust N, Belopolski I, Yuan Z, Bian G, Chang T~R, Zheng H, Strocov V~N, Sanchez D~S, Chang G, Zhang C, Mou D, Wu Y, Huang L, Lee C~C, Huang S~M, Wang B, Bansil A, Jeng H~T, Neupert T, Kaminski A, Lin H, Jia S and Hasan M~Z 2015 {\em Nature Physics\/} {\bf 11} 748--754

\bibitem{Xu2016}
Xu N, Weng H~M, Lv B~Q, Matt C~E, Park J, Bisti F, Strocov V~N, Gawryluk D, Pomjakushina E, Conder K, Plumb N~C, Radovic M, Aut{\`{e}}s G, Yazyev O~V, Fang Z, Dai X, Qian T, Mesot J, Ding H and Shi M 2016 {\em Nature Communications\/} {\bf 7}

\bibitem{Son2013}
Son D~T and Spivak B~Z 2013 {\em Phys. Rev. B\/} {\bf 88}(10) 104412

\bibitem{Zyuzin2012}
Zyuzin A~A and Burkov A~A 2012 {\em Phys. Rev. B\/} {\bf 86}(11) 115133

\bibitem{Sukhachov2020}
Sukhachov P~O and Gorbar E~V 2020 {\em Phys. Rev. B\/} {\bf 102}(1) 014513

\bibitem{Han2020}
Han F, Andrejevic N, Nguyen T, Kozii V, Nguyen Q~T, Hogan T, Ding Z, Pablo-Pedro R, Parjan S, Skinner B, Alatas A, Alp E, Chi S, Fernandez-Baca J, Huang S, Fu L and Li M 2020 {\em Nature Communications\/} {\bf 11}

\bibitem{Arjona2019}
Arjona V, Chernodub M~N and Vozmediano M~A~H 2019 {\em Phys. Rev. B\/} {\bf 99}(23) 235123

\bibitem{Fu2020}
Fu C, Guin S~N, Scaffidi T, Sun Y, Saha R, Watzman S~J, Srivastava A~K, Li G, Schnelle W, Parkin S~S~P, Felser C and Gooth J 2020 {\em Research\/} {\bf 2020} 1--8

\bibitem{Singh2021}
Liu P~F, Li J, Zhang C, Tu X~H, Zhang J, Zhang P, Wang B~T and Singh D~J 2021 {\em Phys. Rev. B\/} {\bf 104}(23) 235422

\bibitem{Yang2016}
Yang S~A 2016 {\em SPIN\/} {\bf 06} 1640003

\bibitem{Mele2018}
Armitage N~P, Mele E~J and Vishwanath A 2018 {\em Rev. Mod. Phys.\/} {\bf 90}(1) 015001

\bibitem{Lv2021}
Lv B~Q, Qian T and Ding H 2021 {\em Rev. Mod. Phys.\/} {\bf 93}(2) 025002

\bibitem{Ruan2016}
Ruan J, Jian S~K, Zhang D, Yao H, Zhang H, Zhang S~C and Xing D 2016 {\em Phys. Rev. Lett.\/} {\bf 116}(22) 226801

\bibitem{Chang2018}
Chang G, Wieder B~J, Schindler F, Sanchez D~S, Belopolski I, Huang S~M, Singh B, Wu D, Chang T~R, Neupert T, Xu S~Y, Lin H and Hasan M~Z 2018 {\em Nature Materials\/} {\bf 17} 978--985

\bibitem{Liu_2019}
Liu J, Fang C and Fu L 2019 {\em Chinese Physics B\/} {\bf 28} 047301

\bibitem{Guo2019}
Guo Y, Lin Z, Zhao J~Q, Lou J and Chen Y 2019 {\em Scientific Reports\/} {\bf 9}

\bibitem{Isobe2016}
Isobe H and Nagaosa N 2016 {\em Phys. Rev. Lett.\/} {\bf 116}(11) 116803

\bibitem{Hao2016}
Hao L and Ting C~S 2016 {\em Phys. Rev. B\/} {\bf 94}(13) 134513

\bibitem{Bader2010}
Bader S and Parkin S 2010 {\em Annual Review of Condensed Matter Physics\/} {\bf 1} 71--88

\bibitem{Li2016}
Li X and Wu X 2016 {\em WIREs Computational Molecular Science\/} {\bf 6} 441--455

\bibitem{Awschalom2007}
Awschalom D~D and Flatt{\'{e}} M~E 2007 {\em Nature Physics\/} {\bf 3} 153--159

\bibitem{Shi2021}
Shi Y, Li L, Cui X, Song T and Liu Z 2021 {\em physica status solidi (RRL) – Rapid Research Letters\/} {\bf 15} 2100115

\bibitem{Meng2021}
Meng W, Zhang X, Liu Y, Wang L, Dai X and Liu G 2021 {\em Applied Surface Science\/} {\bf 540} 148318 ISSN 0169-4332

\bibitem{Li2021}
Li G~G, Xie R~R, Ding L~J, Ji W~X, Li S~S, Zhang C~W, Li P and Wang P~J 2021 {\em Phys. Chem. Chem. Phys.\/} {\bf 23}(21) 12068--12074

\bibitem{He2020}
He T, Zhang X, Liu Y, Dai X, Liu G, Yu Z~M and Yao Y 2020 {\em Phys. Rev. B\/} {\bf 102}(7) 075133

\bibitem{Jia2020}
Jia T, Meng W, Zhang H, Liu C, Dai X, Zhang X and Liu G 2020 {\em Frontiers in Chemistry\/} {\bf 8} ISSN 2296-2646

\bibitem{Wei2022}
Wei X~P, Yang N, Shen J and Tao X 2022 {\em Physica E: Low-dimensional Systems and Nanostructures\/} {\bf 140} 115164 ISSN 1386-9477

\bibitem{You2019}
You J~Y, Chen C, Zhang Z, Sheng X~L, Yang S~A and Su G 2019 {\em Phys. Rev. B\/} {\bf 100}(6) 064408

\bibitem{Zou_2021}
Zou X, Mao N, Li B, Sun W, Huang B, Dai Y and Niu C 2021 {\em New Journal of Physics\/} {\bf 23} 123018

\bibitem{Zhao2022}
Zhao X, Ma F, Guo P~J and Lu Z~Y 2022 {\em Phys. Rev. Res.\/} {\bf 4}(4) 043183

\bibitem{Song2013}
Song Q, Wang B, Deng K, Feng X, Wagner M, Gale J~D, Müllen K and Zhi L 2013 {\em J. Mater. Chem. C\/} {\bf 1}(1) 38--41

\bibitem{Du2017}
Du Q~S, Tang P~D, Huang H~L, Du F~L, Huang K, Xie N~Z, Long S~Y, Li Y~M, Qiu J~S and Huang R~B 2017 {\em Scientific Reports\/} {\bf 7}

\bibitem{galvao2012}
Brunetto G, Autreto P~A~S, Machado L~D, Santos B~I, dos Santos R~P~B and Galvão D~S 2012 {\em The Journal of Physical Chemistry C\/} {\bf 116} 12810--12813

\bibitem{Kresse1993}
Kresse G and Hafner J 1993 {\em Physical Review B\/} {\bf 47} 558--561

\bibitem{Kresse1996}
Kresse G and Furthm\"uller J 1996 {\em Phys. Rev. B\/} {\bf 54}(16) 11169--11186

\bibitem{Perdew1996}
Perdew J~P, Burke K and Ernzerhof M 1996 {\em Physical Review Letters\/} {\bf 77} 3865--3868

\bibitem{Wu2017}
Wu Q, Zhang S, Song H~F, Troyer M and Soluyanov A~A 2018 {\em Computer Physics Communications\/} {\bf 224} 405 -- 416 ISSN 0010-4655

\bibitem{Kochaev2020}
Meftakhutdinov R~M, Sibatov R~T and Kochaev A~I 2020 {\em Journal of Physics: Condensed Matter\/} {\bf 32} 345301

\bibitem{Kremer2023}
Kremer L~F and Baierle R~J 2024 {\em Diamond and Related Materials\/} {\bf 141} 110689

\bibitem{Liu2011}
Liu C~C, Jiang H and Yao Y 2011 {\em Phys. Rev. B\/} {\bf 84}(19) 195430

\bibitem{Soluyanov2011}
Soluyanov A~A and Vanderbilt D 2011 {\em Phys. Rev. B\/} {\bf 83}(23) 235401

\bibitem{Weyl1929}
Weyl H 1929 {\em Proceedings of the National Academy of Sciences\/} {\bf 15} 323--334

\bibitem{Nielsen1983}
Nielsen H and Ninomiya M 1983 {\em Physics Letters B\/} {\bf 130} 389--396 ISSN 0370-2693

\bibitem{Nielsen1981}
Nielsen H and Ninomiya M 1981 {\em Nuclear Physics B\/} {\bf 185} 20--40 ISSN 0550-3213

\bibitem{parkin2021}
Bedoya-Pinto A, Liu D, Tan H, Pandeya A~K, Chang K, Zhang J and Parkin S~S~P 2021 {\em Advanced Materials\/} {\bf 33} 2008634

\bibitem{yan2015}
Sun Y, Wu S~C and Yan B 2015 {\em Phys. Rev. B\/} {\bf 92}(11) 115428

\bibitem{murakami2014}
Okugawa R and Murakami S 2014 {\em Phys. Rev. B\/} {\bf 89}(23) 235315

\bibitem{sarma2018}
Wilson J~H, Pixley J~H, Huse D~A, Refael G and Das~Sarma S 2018 {\em Phys. Rev. B\/} {\bf 97}(23) 235108

\bibitem{Lu2023}
Lu Q, Reddy P~V~S, Jeon H, Mazza A~R, Brahlek M, Wu W, Yang S~A, Cook J, Conner C, Zhang X, Chakraborty A, Yao Y~T, Tien H~J, Tseng C~H, Yang P~Y, Lien S~W, Lin H, Chiang T~C, Vignale G, Li A~P, Chang T~R, Moore R~G and Bian G 2023 {\em arXiv preprint arXiv:2303.02971\/}

\bibitem{Burkov2018}
Burkov A 2018 {\em Annual Review of Condensed Matter Physics\/} {\bf 9} 359--378

\bibitem{Berry1984}
Berry M~V 1984 {\em Proceedings of the Royal Society of London. A. Mathematical and Physical Sciences\/} {\bf 392} 45--57

\end{thebibliography}

\end{document}